\documentclass[a4paper,conference]{IEEEtran}
\usepackage[T1]{fontenc}
\usepackage{color}
\usepackage{float}
\usepackage{mathrsfs}
\usepackage{amsmath}
\usepackage{amssymb}
\usepackage{graphicx}
\usepackage{float} 

\usepackage{CJK}
\usepackage{indentfirst}
\usepackage{amsmath}
\usepackage{cases}
\usepackage{setspace}

\usepackage[unicode=true,
 bookmarks=true,bookmarksnumbered=true,bookmarksopen=true,bookmarksopenlevel=1,
 breaklinks=false,pdfborder={0 0 0},pdfborderstyle={},backref=false,colorlinks=false]
 {hyperref}

\hypersetup{
 pdftitle={Your Title},
 pdfauthor={Your Name},
 pdfpagelayout=OneColumn, 
 pdfnewwindow=true, 
 pdfstartview=XYZ, 
 plainpages=false}
\makeatletter

\floatstyle{ruled}
\newfloat{algorithm}{tbp}{loa}
\providecommand{\algorithmname}{Algorithm}
\floatname{algorithm}{\protect\algorithmname}

\usepackage[caption=false,font=footnotesize]{subfig}
\usepackage{algorithm}
\usepackage{algorithmic}

\usepackage{multirow} %multirow for format of table 
\usepackage{amsmath} 
\usepackage{xcolor}

\allowdisplaybreaks[4]

\ifCLASSOPTIONcompsoc
\usepackage[caption=false,font=normalsize,labelfont=sf,
textfont=sf]{subfig}
\else
\usepackage[caption=false,font=footnotesize]{subfig}
\fi

\usepackage{cite}
\usepackage{bm}
\usepackage{algorithmic}
\usepackage{algorithm}
\usepackage{graphicx}
\IEEEoverridecommandlockouts

\usepackage{lettrine}

\usepackage{geometry}
\usepackage{subfig}
\makeatother

\begin{document}

\title{\textcolor{black}{
    Trusted Routing for Blockchain-Enabled  Low-Altitude Intelligent Networks}}   
\author{
    \IEEEauthorblockN{Sijie He$^{\dagger}$, 
        Ziye Jia$^{\dagger}$,
        Qiuming Zhu$^{\dagger}$, 
        Fuhui Zhou$^{\dagger}$,
        % Yilu Cao$^{\dagger}$, 
        % Yang Yang$^{\ddagger }$, 
        and Qihui Wu$^{\dagger}$\\}
        \IEEEauthorblockA{$^{\dagger}$The Key Laboratory of Dynamic Cognitive System of 
         Electromagnetic Spectrum Space, Ministry of Industry and\\
          Information Technology, 
         Nanjing University of Aeronautics and Astronautics, Nanjing, Jiangsu, 211106, China\\
          \{hesijie, jiaziye, zhuqiuming, zhoufuhui, wuqihui\}@nuaa.edu.cn}
        
\thanks{{
This work was supported in part by the Natural Science Foundation on 
Frontier Leading Technology Basic Research Project of Jiangsu under Grant BK20222001, 
in part by National Natural Science Foundation of China under Grant 62301251, 
in part by the Aeronautical Science Foundation of China 2023Z071052007, 
and in part by the Young Elite Scientists Sponsorship Program by CAST 2023QNRC001.
}} 
}
        
\maketitle
% \pagestyle{fancy}
% \fancyhf{}
% \fancyhead[R]{\fontsize{7}{9}\selectfont \thepage}
% \renewcommand{\headrulewidth}{0pt} 
% \renewcommand{\footrulewidth}{0pt}

\pagestyle{empty} 

\thispagestyle{empty}

\begin{abstract}
    Due to the scalability and portability, the low-altitude intelligent networks (LAINs)
    are essential in various fields such as surveillance and disaster rescue. 
    However, in LAINs, unmanned aerial vehicles (UAVs) are characterized by 
    the distributed topology and high dynamic mobility, and vulnerable to security threats,
    which may degrade the routing performance for data transmission. 
    Hence, how to ensure the routing stability and security of LAINs is a challenge.
    In this paper, we focus on the routing process in LAINs with multiple UAV clusters 
    and propose the blockchain-enabled zero-trust architecture to manage the joining and exiting of UAVs. 
    Furthermore, we formulate the routing problem to minimize the end-to-end (E2E)
    delay, which is an integer linear programming and intractable to solve.
    Therefore, considering the distribution of LAINs, 
    we reformulate the routing problem into a decentralized partially observable Markov decision process.
    With the proposed soft hierarchical experience replay buffer, 
    the multi-agent double deep Q-network based adaptive routing algorithm is designed.  
    Finally, simulations are conducted and numerical results show that the total 
    E2E delay of the proposed mechanism decreases by 22.38\% than the benchmark on average.

\end{abstract}

\newcommand{\CLASSINPUTtoptextmargin}{0.8in}

\newcommand{\CLASSINPUTbottomtextmargin}{1in}
% The trust mechanism employ blockchain as a trusted ledger to record status of UAVs. 
% due to the diversity of UAV network deployment scenarios and the features of high- mobility and dynamic topology.
% that provide verifiable and traceable records of interactions {\cite{Trusted_Blockchain}}. 
% Additionally, in dense UAV networks or environments with obstacles, low delay is crucial for collision avoidance systems. 
% Quick data exchange ensures that UAVs can detect and respond to potential collisions in time.
% At the same time, security is a major factor considered when a system is designed, and invariably, the efficiency of UAV network data routing is heavily influenced by security.
% For example, in {\cite{MARL-o-Manage-1}}, to address the multi-UAV cooperative task scheduling, 
% the author proposes the clustering-based multi-agent deep deterministic plicy gradient algorithms 
% which leverages dynamic UAV clustering to partition UAVs into clusters, each managed by a cluster head UAV, 
% facilitating a distributed-centralized control approach. 
\section{Introduction}
\lettrine[lines=2]{A}{s} key components of the six generation communication networks, 
the low-altitude intelligent network (LAINs) are widely applied to multiple tasks, 
such as disaster rescue and real-time monitoring {\cite{10418158, 10599389, 11006480}}. 
In these applications, the data generated from sensor devices (SDs) are required to be 
relayed to the remote ground base stations (BSs) by unmanned aerial vehicles (UAVs){\cite{10899883}}.
Particularly, UAVs act as aerial relays and cooperatively accomplish the data collection
and transmission, providing low-cost, flexible, and versatile services.
In LAINs, routing is a significant issue for data transmission  {\cite{Routing_UAV_Survey, He_Routing, 10574195}}. 

However, since UAVs in LAINs are characterized by the complex application
environment, high mobility and distributed topology, they are vulnerable 
to security threats and unreliable, i.e., attacks and node failures.
Thus, the availability of communication links is susceptible, 
which leads to the reduction of routing performances. 
Hence, it is significant to efficiently manage the mobility of UAVs in the zero-trust environment.
There exist a couple of works related to the management of UAVs.
For instance, in {\cite{GCS_1}}, the ground control station (GCS) is responsible 
for managing the connection among all UAVs, through the status message sent by UAVs.
In {\cite{GCS_2}}, the authors present that as the center controller and manager, 
the GCS receives the data transmitted by UAVs, which is suffered from long distances, 
undulating terrains or other interferences.
Authors in {\cite{GCS_3}} present that the GCS remotely controls UAVs by sending controlling information, 
in which the relay UAV is limited within the communication range of the GCS, 
restricting the scope of operations.
Moreover, the above works rely on a center controller, which is not resilient 
to fault tolerance and may be susceptible to tampering.
Hence, how to guarantee the reliability and security of routing remains challenging
in the distributed and zero-trust network.

Meanwhile, the transmission delay is a key quality of service,
since low delay can significantly improve the timeliness and reliability for 
various emergency applications, such as the surveillance information transmission {\cite{ Wang_security}}.
Therefore, it is necessary to  design an adaptive and 
dynamic algorithm to enable timely and reliable routing 
in the varying network topology.
To tackle this issue, the multi-agent reinforcement learning (MARL) can be applied {\cite{10638237}}.
Specifically, there exist a couple of works focusing on MARL-based routing problems in dynamic scenarios. 
For example, in {\cite{MARL-o-Manage-4}}, the authors propose a value decomposition network based MARL algorithm to 
minimize the end-to-end (E2E) delay of packet routing within dynamic aerial and terrestrial hybrid networks.
{\cite{MARL-o-Manage-5}} formulates the packet routing as a max-min problem using the Lagrange method, 
% while prioritizing fast communication and meeting energy efficiency and packet loss requirements, 
and proposes a constrained MARL dynamic routing algorithm to balance the objective improvement and constraint satisfaction.
Authors in {\cite{MARL-o-Manage-6}} establish the mean-field enhanced heterogeneous MARL framework
to optimize the communication energy efficiency during routing.
The above studys illustrate that the MARL can solve the routing problem effectively.
However, these works do not consider the mobility management of UAVs in the zero-trust environment.

To deal with the above challenges, in this paper, the routing process is depicted 
in the zero-trust LAINs with multiple UAV clusters, considering the dynamic joining and exiting of UAVs.   
Meanwhile, to improve the network reliability and security, the blockchain technique with a distributed ledger is 
introduced to manage the mobility of nodes and avoid being tempered with, by leveraging the smart contract. 
Further, a couple of UAVs with powerful capabilities are selected to construct the decentralized controller in the blockchain.
In light of the constructed LAIN, 
the routing problem is formulated to minimize the total E2E delay, 
which is an integer linear programming (ILP) and intractable to solve. 
Besides, it is tricky to obtain the global information due to decentralized UAVs.
Hence, we reformulate the routing problem into a decentralized partially observable Markov decision process (Dec-POMDP).
To improve the efficiency of learning and training, 
we propose the multi-agent double deep Q-network (MADDQN)-based adaptive routing approach 
with the designed soft hierarchical experience replay buffer (SHERB). 
Finally, numerical simulations are conducted to verify the performance of the proposed algorithm.

\vspace{0.5cm}
\section{System Model and Problem Formulation\label{Sec:System-model}}
\vspace{0.1cm}
\subsection{Network Model}
\begin{figure}[t]
\vspace{-0.1cm}
    \centering
    \includegraphics[width=0.9\linewidth]{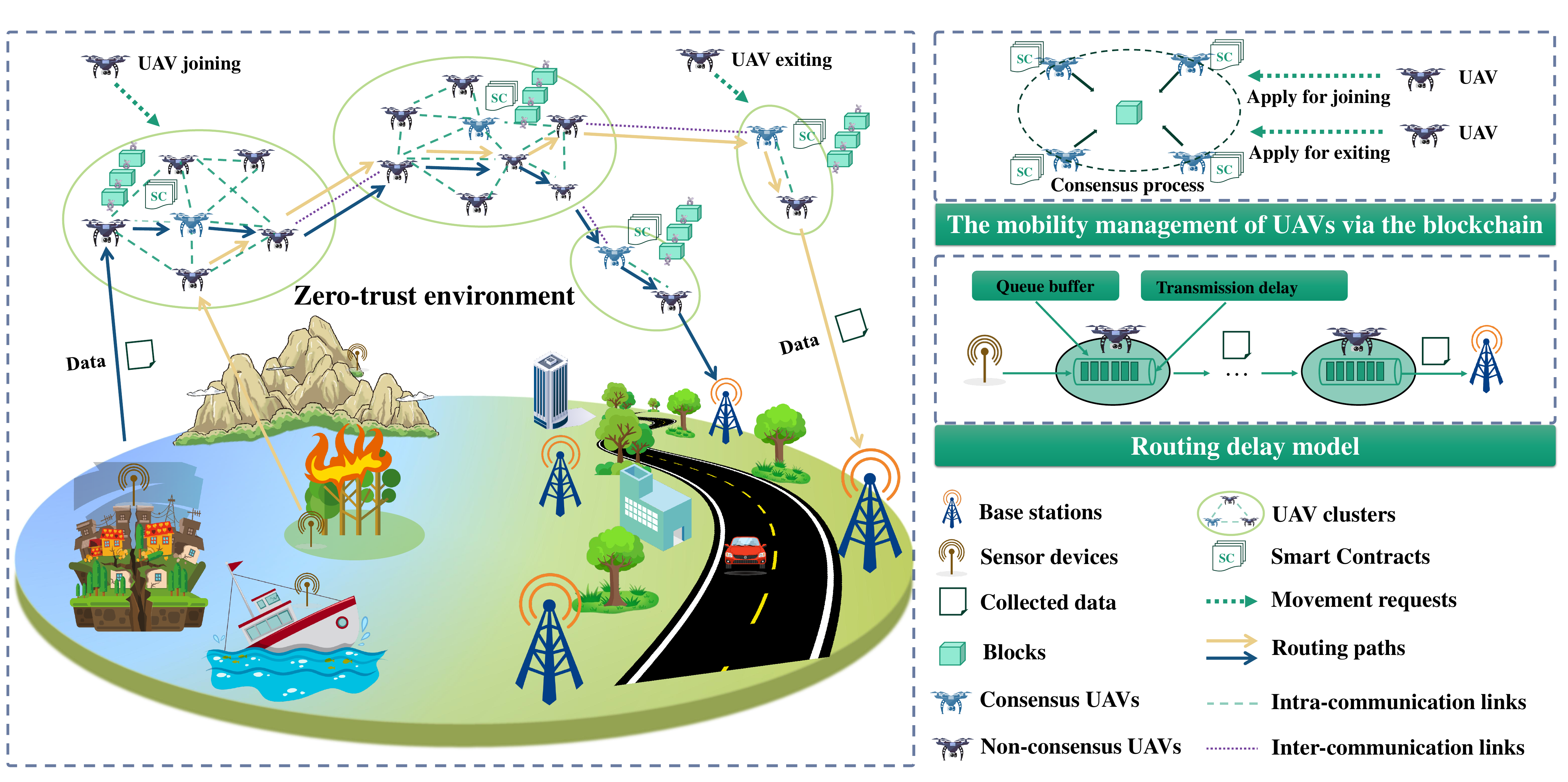}
    % \small % 这里设置字体为小一号，可按需调整
    \caption{\label{fig:Network_of_UAV} Routing scenario in zero-trust LAINs with the mobility management of UAVs via the blockchain technique. 
    } 
\vspace{-0.1cm}
\end{figure}

The routing process in zero-trust LAINs is show in Fig. \ref{fig:Network_of_UAV}, 
including $N$ nodes, and each node holds a unique identity. 
Specifically, when there exist UAVs applying to joining or exiting, 
the identities are authenticated and managed by smart contracts via the blockchain technique.
Moreover, the periodical authentication is performed on all UAVs for the persistent verification.
In detail, the ground layer consists of $B$ BSs and $I$ SDs.
%  where the SDs are far away from BSs. 
The air layer is composed of $U$ UAVs. 
$\mathcal{B}\! =\!\{1,...,b,...,B\}$, $\mathcal{I}\! =\!\{1,...,i,...,I\}$, 
and $\mathcal{U}\! =\!\{1,...,u,...,U\}$ denote the sets of 
BSs, SDs, and UAVs, respectively. 
Besides, UAV set $\mathcal{U}=\mathcal{U}_\mathtt{s} \cup \mathcal{U}_\mathtt{r} \cup \mathcal{U}_\mathtt{d}$ 
is divided based on clusters, in which $\mathcal{U}_\mathtt{s}$, $\mathcal{U}_\mathtt{r}$, 
and $\mathcal{U}_\mathtt{d}$ denote the data collection, relay forwarding, and downlinking clusters, respectively.
In each cluster, the UAV with the most energy and computing power is selected as the cluster head.
$\mathcal{E}\!=\!\mathcal{E}_{iu}\!+\!\mathcal{E}_{uu}\!+\!\mathcal{E}_{ub}$ indicates the set of communication 
link status among all nodes, where $e_{iu}\!\in\!\mathcal{E}_{iu}$, $e_{ub}\!\in\!\mathcal{E}_{ub}$, 
and $e_{u u }\!\in\!\mathcal{E}_{uu}$ represent the connection of links
between the SD and UAV, the BS and UAV, and the UAV and UAV, respectively.
$e_{nm}\in \mathcal{E}$ indicates whether there exists a communicable link 
between nodes $n\in \mathcal{I} \cup \mathcal{U}$ and $m\in \mathcal{U} \cup \mathcal{B}$. 
Specifically, $e_{nm}=1$ indicates the link is effective, 
and $e_{nm}=0$ denotes there exist no direct links. 

In particular, the time period is divided into $\mathsf{T}$ steps. 
% and the length of each time step is $\tau$
The set of time steps is represented as $T \!=\! \{1,...,t,...,\mathsf{T}\}$.      
At the beginning of time step $t$, demand $r$ is transmitted from SD $i\in \mathcal{I}$ to 
destination BS $b\in \mathcal{B}$, denoted by 
$\varUpsilon^r\!=\!\{\mathtt{s}^r,\mathcal{L}^r,\mathtt{d}^r,\mathcal{T}^r_{max}\}$,  
in which the source is $\mathtt{s}^r\!=\!i$, and the destination is $\mathtt{d}^r\!=\!b$.
Here, the size of demand $r$ is $\mathcal{L}^r$ (in bit), 
and $\mathcal{T}^r_{max}$ is the maximum delay tolerance of the demand transmission.
Further, demands are uploaded from SD $i \!\in\! \mathcal{I}$ to UAV $u\!\in\! \mathcal{U}_\mathtt{s}$,  
relayed by UAV $u\! \in\! \mathcal{U}_\mathtt{r}$, and downloaded 
from UAV $u\!\in\!\mathcal{U}_\mathtt{d}$ to BS $b \!\in\! \mathcal{B}$. 
$\mathcal{P}_{ib}^{r}\!=\!{(e_{iu},\cdots,e_{ub})}$ 
indicates a completed routing path for transmitting demand $r$. 
% from source SD $i\!\in\!\mathcal{I}$ to corresponding destination BS $b\! \in\!\mathcal{B}$. 

The coordinates of SD $i$ 
and BS  $b$ remain fixed and are ${\varTheta}_{i}\!=\!{(x_{i},y_{i},0)}$ and 
${\varTheta}_{b}\!=\!{(x_{b},y_{b},0)}$, respectively. 
The location of UAV $u$  is indicated by 
${\varTheta}_{u}(t)={(x_{u}(t),y_{u}(t),z_{u}(t))}$ in three-dimensional Cartesian coordinates at time step $t$. 
% In one time step, the UAV is quasi-static. 
% The velocity in $x$, $y$, and $z$ directions of UAV $u$ is 
% represented by ${{{\bm{\nu}}}}_{u(t)}={(\nu^{x}_{u(t)},\nu^{y}_{u(t)},\nu^{z}_{u(t)})}$, 
% and the velocities of all UAVs are denoted as
%  $\bm{V(t)}=\{\bm{\nu}_{1(t)},\cdots,\bm{\nu}_{u(t)},\cdots,\bm{\nu}_{U(t)}\}$ at time step $t$. 
% Further, $\bm{\nu}_{i}=(\bm{\nu}_{i}(1),\bm{\nu}_{i}(2),\cdots,\bm{\nu}_{i}(T))$ indicates the velocities sequence of UAV $i\in\mathcal{U}$. 
Additionally, the Euclidean distance between nodes $n \in \mathcal{I}\cup\mathcal{U}$ and $m \in \mathcal{U}\cup\mathcal{B}$
is indicated by $d_{n  m}(t)$ at time step $t$, i.e., 
\begin{equation}{\label{distance}}
    \begin{aligned}
    &d_{n  m}(t) \!= \!\\
    &\sqrt{(x_{n}(t)\!-\! x_{m}(t))^2\!+\!(y_{n}(t)\!-\!y_{m}(t))^2\!+\!(z_{n}(t)\!-\!z_{m}(t))^2}.
    \end{aligned}
\end{equation}
In particular, the distance between UAVs should satisfy
\begin{equation}{\label{distance_min}}
    d_{min} \leqslant d_{n  m}(t), \forall n ,m \in\mathcal{U}, n\neq m, t\in T.
\end{equation}
Here, $d_{min}$ represents the safe distance to avoid collisions among UAVs.
At time step $t$, the set of connected UAVs for UAV $u$ is denoted as $\Gamma_{u}(t)$,   
and distance $d_{u\kappa }(t)$ between UAV $u$ and UAV $\kappa \in \Gamma_{u}(t) $ satisfies 
\begin{equation}{\label{distance_max}}
    d_{u \kappa}(t)\leqslant d_{u,max},\forall \kappa \in \Gamma_{u}(t), u\in \mathcal{U}, t \in T,
\end{equation}
in which $d_{u,max}$ is the maximum communication distance of UAV $u$.

\vspace{0.2cm}
\subsection{Blockchain Model}
To manage the mobility of UAVs, a lightweight blockchain model is proposed.
% In detail, when there exists a new UAV applies to join the LAIN, 
% which is required for authentication from the blockchain. 
Additionally, to maintain the sustainability and reliability of routing, 
the joining and exiting of UAVs need to be recorded and updated via the blockchain.

\subsubsection{Role Selection}
In the lightweight blockchain, the set $\mathcal{U}$ of UAVs is divided into the sets of full nodes 
$\mathcal{U}_f= \{1, 2,  \cdots, u_f \}$ and light nodes $\mathcal{U}_l= \{1, 2, \cdots, u_l \}$, respectively.
In detail, as shown in Fig. {\ref{fig:Network_of_UAV}}, 
cluster head UAVs are selected as full nodes due to the sufficient resources,
while the other UAVs act as light nodes in LAINs. 
In particular, the full nodes have the full blockchain ledger and are responsible for the blockchain consensus, 
including broadcasting and verification of transactions. 
The light nodes can only store the header of blocks and are in charge of generating local transactions and relaying transactions from other UAVs.

\subsubsection{PBFT-based Consensus Process}
We apply the practical Byzantine fault tolerance (PBFT) method, 
which can ensure the success of the consensus process when the number of failure nodes is less than one third. 
% However, , the random selection of the primary node may cause the entire consensus process to fail when it is the malicious node. 
% Moreover, due to the larger load on primary nodes, selecting capable nodes is significant for the effectiveness and reliability of the consensus. 
For the effectiveness and reliability of the PBFT consensus, 
% due to the larger load on the primary node, 
it is significant to select consensus UAVs with the most abundant resources as the primary node, 
while other consensus UAVs operate as non-primary (replica) nodes. 

\subsubsection{Transaction Generation and Broadcasting}
The transactions waiting for the consensus are generated periodically and include the status of UAVs, 
i.e., the queue buffer, location, and information of neighboring nodes.
Then, all generated transactions are broadcasted to the entire network  via the Gossip protocol for verifying. 
After verification, the transactions are added to the transaction pool of the blockchain and then synchronized to the 
entire blockchain. Once a consensus is reached, the data transaction are recorded in the blockchain with tamper-proof.

\subsection{Delay Model}
\vspace{0.05cm}
As shown in Fig. {\ref{fig:Network_of_UAV}}, in LAINs, the  E2E  routing delay consists of the total
transmission delay on the multi-hop path from the source UAV to destination UAV.
Since the delay is constrained by data transmission rate $G$, it is essential to 
analyze the channel characteristics of both ground-to-air and air-to-air wireless communication links.
% For the simplicity and clarity, the Definition 1 is first presented.
% Definition 1: $X_{n m}(t)$ denotes the set of variants, detailed as $X\!=\!\{G,L,B,P^{tr},\sigma,\eta \}$. 
% Besides, $X_{n m}(t)\!=\!X_{i u}(t) (i\in \mathcal{I},u\in \mathcal{U})$, 
% $X_{nm}(t)\!=\!X_{u \kappa}(t) (u \! \in \! \mathcal{U}, \kappa \!\in\! \Gamma_{u(t)})$, and
% $X_{n m}(t)\!=\!X_{u b}(t) (u\!\in\! \mathcal{U}, b\!\in\! \mathcal{B})$ indicates the variant from SD $i$ to UAV $u$, 
% UAV $u$ to UAV $\kappa$, and UAV $u$ to BS $b$, respectively. 

Regard to demand $r$, binary variable $\zeta_{n}^{r}(t) \!\in\! \{0,1\}$ clarifies 
the transmission status, i.e.,
\begin{equation}{\label{zeta-demands}}
    \zeta_{n}^{r}(t)\!=\!
    \left\{
        \begin{aligned}
            &1, \!\text{ if }\!  r \! \text { is on node }  n  \in \mathcal{I}\cup \mathcal{U}\text { at time step }\! t,\\
            &0, \!\text{ otherwise}.
        \end{aligned}
    \right.
\end{equation}
Besides,  binary variable $\eta_{n m}^{r}(t) \!\in\! \{0,1\}$ denotes whether demand $r$ 
passes  link $e_{n m}$ ($n  \in \mathcal{I}\cup \mathcal{U},m  \in \mathcal{U}\cup \mathcal{B}$), i.e.,
\begin{equation}{\label{eta-link}}
    \eta_{n m}^{r}(t)\!=\!
    \left\{
        \begin{aligned}
            &1, \!\text{ if }\!  r \! \text { is transmitted via }\! e_{n m}\! \text { at time step }\! t,\\
            &0, \!\text{ otherwise}.
        \end{aligned}
    \right.
\end{equation}
Further, $L_{n m}(t)$ is the path loss between nodes $n \in \mathcal{I} \cup \mathcal{U}$ and $m \in \mathcal{U} \cup \mathcal{B}$
and remains constant within time step $t$, i.e., 
\begin{equation}{\label{PLAG}}
    \begin{aligned}
        &L_{n m}(t)=20\log \left(\frac{4\pi  \lambda   }c d_{n m}(t)\right)\\
        &+\omega\left[{\Pr}_{n m}(t) \left(\eta_{n m}^\mathrm{LoS}(t)-\eta_{nm}^\mathrm{NLoS}(t)\right) +\eta_{n m}^\mathrm{NLoS}(t)\right],
    \end{aligned}
\end{equation} 
where $\eta_{n m}^\mathrm{LoS}(t)$ and $\eta_{nm}^\mathrm{NLoS}(t)$  
represent the additional path losses under LoS and non-LoS (NLoS) propagations, respectively. 
When $\omega=1$, $L_{n m}(t)$ is  the path loss  from $n\in \mathcal{I}$ to $m\in \mathcal{U}$ or $n\in \mathcal{U}$ to $m\in \mathcal{B}$.
While $\omega=0$, $L_{n m}(t)$ indicates the path loss between UAVs. 
$\lambda  $ represents the carrier frequency, and $c$ is the speed of light.
Besides, ${\Pr}_{n m}(t)$ indicates the probability that  there exist line of sight (LoS) links between SDs/BSs and UAVs, i.e.,
\begin{equation}{\label{P}}
    {\Pr}_{n m}(t)\!=\!\frac1{1\!+\!\varrho _1\!\exp\left\{\!-\!\varrho_2 \!\left[ \! \frac{180}{\pi}\!\arctan \! \left( \! \frac{h_{n m}(t)}{xy_{n  m}(t)} \! \right)\!-\!\varrho _1 \! \right] \!\right\}},
\end{equation} 
in which $\varrho_1$ and $\varrho_2$ are the constant parameters {\cite{channel_model_PL}}.

Since LAINs are characterized by high mobility and unstable data traffic fluctuations,
we introduce a queue buffer for each node $n \in \mathcal{I} \cup \mathcal{U}$ to alleviate the network congestion.
In particular, for node $n$, the amount of demands received at time step $t\!-\!1$ is $\varepsilon_{n}^{re}(t\!-\!1)$. 
At time step $t$, the queued demand set, queue length, maximum queue capacity, 
and the number of transmitted demands of node $n$ are denoted as 
$\mathcal{C}_{n}(t)=\{r , \zeta_{n}^{r}(t)\!=\!1, \forall r \!\in\! R\}$, $C_{n}(t)$, $C_{n}^{max}$, and $\varepsilon_{n}^{tr}(t)$, respectively.
In detail,
\begin{align}{\label{re_tr}}
    \left\{\begin{aligned}
    &\varepsilon_{n}^{re}(t\!-\!1)\!=\!\sum_{r\in R} \sum_{{m}  \in \mathcal{I}  \cup \mathcal{U} } \eta^r_{{m}n}(t\!-\!1), 0\!\leq\! \varepsilon_{n}^{re}(t\!-\!1)\!\leq\!C_{n}^{max},\\
    &C_{n}(t)=\sum_{r\in R}\zeta_{n}^{r}(t), 0\leq C_{n}(t)\!\leq \!C_{n}^{max}, C_{n}(t)\!=\!\left\lvert \mathcal{C}_{n}(t)\right\rvert, \\
    &\varepsilon_{n}^{tr}(t)\!=\!\sum_{r\in \mathcal{C}_{n}(t)} \sum_{{m}  \in  \mathcal{U}\cup \mathcal{B}  } \eta^r_{n {m}}(t),\varepsilon_{n}^{tr}(t)\!=\! C_{n}(t).
\end{aligned}\right.
\end{align}
% 无人机采用并行传输的方式传输请求,我们采用自适应带宽分配方案,缩小统一时隙不同数据包传输时延差距。
Here, the UAV leverages the parallel transmission for demands, 
indicating that UAVs forward all demands received from the last time step.
Besides, the adaptive channel bandwidth allocation scheme is designed 
to narrow the transmission delay gap of different demands in one time step.
Hence, at time step $t$, the allocated bandwidth $B^r_n(t)$ is calculated based on the size of demand $r \in \mathcal{C}_{n}(t)$, i.e.,
\begin{equation}{\label{equ:BW}}
    B^r_n(t)\!= \!\frac{\mathcal{L}^{r}}{\sum\limits_{k  \in \mathcal{C}_{n}\!(t)}\!\mathcal{L}^{k}} B_n(t), 
    \forall n \!\in\! \mathcal{I} \! \cup \!\mathcal{U},r \!\in \!\mathcal{C}_{n}(t), t \!\in \!T,
\end{equation}
where $ B_n(t)$ is the total bandwidth of node $n$ at time step $t$.
Based on Shannon theory, at time step $t$, transmission rate $G^r_{n m}\!(t)$ for demand $r$ from nodes $n\!\in\! \mathcal{I} \! \cup \!\mathcal{U}$ 
to $m\!\in\! \mathcal{U} \! \cup \!\mathcal{B}$ is 
\begin{equation}{\label{Channel-Rate}}  
    G_{n m}^r(t)\!=\!B^r_n(t)\log_{2}\!\left(\!1\!+\!
    \frac{P^{tr}_{n m}(t) \cdot 10^\frac{-{L_{n m}(t)}}{10}}{\sigma^{2}_{n m}(t)}\!\right)\!, r \in \mathcal{C}_{n}(t),
\end{equation}
where $P^{tr}_{n m}(t)$ and $\sigma^{2}_{n m}(t)$
indicate the transmission power and noise power between nodes $n$ and $m$, respectively. 
Then, the transmission delay for demands from nodes $n$ to $m$ is
\begin{equation}{\label{trans-delay1}}
    \begin{aligned}
         \mathcal{T}^{tr,r}_{n  m}(t) = & \max_{r \in R}  \frac{\mathcal{L}^{r}}{G^r_{n  m}(t)} \eta^{r}_{n  m}(t),\\
     &\forall r \in  R, n  \in  \mathcal{I}\cup \mathcal{U}, m  \in  \mathcal{U}\cup\mathcal{B}, t \in  T.
    \end{aligned}
\end{equation}
When the routing path $\mathcal{P}^{r}_{ib}$ for transmitting demand $r$ 
from source SD $i$ to destination BS $b$ 
is determined, E2E delay $\mathcal{T}^{r}$ can 
be calculated as 
% Therefore, the E2E delay from source UAV $\mathtt{s} \in \mathcal{U}$  to destination UAV $\mathtt{d}\in \mathcal{U}$ is indicated as 
\begin{equation}
    \mathcal{T}^{r}\!=\!\underset{t\in T}{\sum} \underset{{e_{n m}\in \mathcal{P}_{ib}^r}}{\sum}\mathcal{T}^{tr,r}_{n  m}(t), \forall r\!\in\! R, n \!\in \!\mathcal{I}\cup \mathcal{U}, m\! \in\! \mathcal{U}\cup\mathcal{B}.
\end{equation}
\vspace{0.1cm}
\subsection{Problem Formulation}

The objective is to minimize the total E2E delay of LAINs with the mobility of nodes, 
and the corresponding optimization problem is formulated as
% In this article, our objective is to achieve efficient transmission during 
% routing by balancing the distance, energy consumption, and delay. 
% Therefore, we consider jointly optimizing the distance 
% factor $f_1$, energy factor $f_2$ and delay factor $f_3$ of the UAV nodes.
% In detail, the joint optimization function is defined as follows
\begin{equation}{\label{optimal}}
    \begin{aligned}
    \mathscr{P}0:\;&\underset{{\boldsymbol{\eta,\zeta}}}{\textrm{min}}\;
    \mathcal{T}^{r}\\
       \textrm{s.t.}\;
        &\text{(\ref{distance_min}), (\ref{distance_max}), (\ref{zeta-demands}), (\ref{eta-link}), (\ref{re_tr})},\\
        &\sum_{m  \in \mathcal{U} \cup \mathcal{B} }\!\eta^{r}_{n m}(t)\!=\!1, \forall r \!\in\! R, t \!\in\! T,e_{n m}\! \in \mathcal{E},\\
        &\forall  n\in \mathcal{I} \cup\mathcal{U}, m \in \mathcal{U}\cup \mathcal{B}, 
    \end{aligned}
\end{equation}
where $\boldsymbol{\eta}=\{\eta_{n m}^{r}(t),\forall t\in T, e_{n m}\in \mathcal{E}, 
n \in \mathcal{I} \cup \mathcal{U},m \in \mathcal{U}\cup \mathcal{B}, r \in R\}$, 
and $\boldsymbol{\zeta}=\{\zeta_{n}^{r}(t),\forall t\in T, n \in \mathcal{I} \cup \mathcal{U}, r \in R\}$.
% in which $n,m \notin \mathcal{U}_\mathtt{f}$ represents that demand $r\in R$ avoids being transmitted by malicious UAVs in the network. 
Besides, the demand $r$ from node $ n \!\in\! \mathcal{I} \cup \mathcal{U}$ can 
only be received by one another node $m \!\in \!\mathcal{U}\cup \mathcal{B}$.
It is observed that $\mathscr{P}0$ is in the form of ILP and 
NP-hard to deal with {\cite{Lancia2018}}. 
% Therefore, in the next part we propose the feasible solutions. 

% Therefore, in the following section, we design the efficient routing algorithm based on the MADDQN.
% Firstly, the distance factor $f_1$ is expressed as
% \begin{equation}
%     f_{1}=\chi \frac{\sqrt{(x_{j(t)}-x_{d(t)})^2,(y_{j(t)}-y_{d(t)})^2,
%     (z_{j(t)}-z_{d(t)})^2}}{\sqrt{(x_{i(t)}-x_{d(t)})^2,(y_{i(t)}-y_{d(t)})^2,
%     (z_{i(t)}-z_{d(t)})^2}}\eta_{i(t)j(t)}^{p_{i(t)}^{m}}
% \end{equation}
% where $\chi\in \{1,-1\}$, and $\chi$ is negative when the next hop UAV $j(t)$ is 
% further away from the destination than UAV $i(t)$ and vice versa. 
% Secondly, the energy factor $f_2$ of UAV $j(t)$ is defined as 
% the ratio of the residual energy $E_{j(t)}^{res}$ of the UAV 
% to the initial energy $E_{j(t)}^{init}$, i.e.,
% \begin{equation}
%     f_{2}= \frac{E_{j(t)}^{res}}{E_{j(t)}^{init}}\eta_{i(t)j(t)}^{p_{i(t)}^{m}},
% \end{equation}
% in which a larger $f_3$ indicates the lower energy consumption of UAV ${j(t)}$.
% Thirdly, $f_3$ represents the delay factor. Then, we introduce ${T}_{base}$ 
% as a base to standardize and normalize the value of ${T}_{i(t) j(t)}$. 
% The delay between UAV ${i(t)}$ and UAV ${j(t)}$ can be expressed as follows
% \begin{equation}
%     f_3=\frac{{T}_{base}-{T}_{i(t) j(t)}}{{T}_{base}}\eta_{i(t)j(t)}^{p_{i(t)}^{m}}.
% \end{equation}

\vspace{0.5cm}
\section{Problem Formulation and Algorithm Design}{\label{Sec:GATDRL}}
\vspace{0.1cm}
\subsection{Dec-POMDP based Reformulation}
\vspace{0.1cm}
Since each UAV has a local observation rather than global observation, 
$\mathscr{P}0$ is reformulated as a Dec-POMDP, to cater for the dynamically changing network environment. 
Each UAV in LAINs is regarded as an independent agent and makes its own routing decision 
that sends a demand to an alternative next UAV. Hence, the set of all agents is equal to UAV set $ \mathcal{U}$.
At each time step $t$, agent $u\in \mathcal{U}$ observes local state $o_{u}(t)$ of the environment 
and executes action $a_{u}(t)$ according to the observable state. 
Then, agent $u$ receives an immediate reward $\mathcal{R}_{u}(t+1)$, 
and the environment is transfered to next observation state $o_{u}({t+1})$.
The tuple $\left\langle o_{u}(t),a_{u}(t),\mathcal{R}_{u}(t),o_{u}({t+1}),f_{u}(t)\right\rangle$ 
indicates the transition experience of agent $u$ and is explained in detail.
\begin{itemize}
    % The goal of each drone agent is to efficiently route carried data packets (with different destination nodes)  to their destinations through collaboration.
    \item State space $\mathcal{S}$: At time step $t$, the observable state $o_{u}(t)$ of agent $u$
    includes the available information of the current UAV and neighboring UAVs, detailed as
    \begin{equation}
        \begin{aligned}
        o_{u}(t)\!=\!\{\Theta_{u}(t), \mathcal{C}_{u}(t), \Theta_{\kappa}(t),\mathcal{C}_{\kappa}(t)\}, 
        \kappa\!\in\!\Gamma_{u}(t).
        \end{aligned}
    \end{equation}
   Wherein, $\Theta_{u}(t)$ and $\Theta_{k}(t)$ indicate the locations of UAV $u$ and neighbor UAV $\kappa$, respectively. 
   $\mathcal{C}_{u}(t)$ and $\mathcal{C}_{\kappa}(t)$ are the set of queued demands on UAVs $u$ and $\kappa$, respectively.
    The queued demand set includes the information tuple $\varUpsilon^r$ of each demand $r$.
   The observations of all agents are aggregated into joint state $\bm{s}(t)$ in time step $t$, 
   denoted as $\bm{s}(t)=\{o_{u}(t), u\in \mathcal{U}\}$,
   and the state space is indicated as $\mathcal{S} = \{\bm{s}(t)| t\in T\}$.
% It is noted that the queue length of UAVs is empty at the beginning of each time step, 应该加上的，因为有优化空间，根据时延先后顺序
    \item Action sapce $\mathcal{A}$: Agent $u$ makes decisions for each carried demand independently, 
    and the set $a_{u}(t)= \{a_{u}^r(t), r \!\in \!\mathcal{C}_{u}(t)\}$ represents the actions 
    for demand $r \!\in \!\mathcal{C}_{u}(t)$ at time step $t$.
    Wherein, each sub-action $a_{u}^r(t)$ denotes the next-hop neighboring node 
    selected by UAV $u$ to relay demand $r$, i.e.,
    \begin{equation}
        \begin{cases}
            a_{u}^r(t) \in \{\Gamma_{u}(t)\},u \in \mathcal{U}_\mathtt{s} \cup \mathcal{U}_\mathtt{r},\\
            a_{u}^r(t) \in \{ \Gamma_{u}(t), Z _u(t)\},  u \in \mathcal{U}_\mathtt{d},
        \end{cases}    
    \end{equation}
    where $Z _u(t)$ is the set of BSs connected with UAV $u \in \mathcal{U}_\mathtt{d}$ at time step $t$. 
    If the destination BS $b$ of demand $r \!\in\! \mathcal{C}_u(t)$ connects to UAV $u \!\in\!\mathcal{U}_\mathtt{d}$, 
    the demand is directly transmitted to destination $b$ by UAV $u$.
    On the contrary, the demand is relayed by neighboring UAVs.
    The actions of all agents are aggregated as $\bm{a}(t)=\{a_{u}(t), u\in \mathcal{U}\}$ in time step $t$,
    and the action space is $\mathcal{A} = \{\bm{a}(t)| t\in T\}$.
    \item Reward $\mathcal{R}$:  $\mathcal{R}_u(t)\!=\!\{\mathcal{R}_{u}^r(t)|r \!\in\! \mathcal{C}_{u}(t)\}$, 
    where $\mathcal{R}_{u}^r(t)$ is the reward that agent $u$ obtains after transmitting demand $r$ to neighboring node 
$\kappa$ at time step $t$, i.e.,
% \begin{equation}
%     \mathcal{R}_{u}(t)=\{\mathcal{R}_{u}^r(t), \forall r \in R \},
% \end{equation}
% and 
\begin{equation}{\label{equ:reward}}
    \mathcal{R}^{r}_{u}(t)=\frac{1}{\mathcal{T}^{tr,r}_{u\kappa}(t)} \eta_{u \kappa}^r(t), \forall r \in \mathcal{C}_u(t), \kappa \in \Gamma_{u}(t).
\end{equation}
    \item Transition flag $\bm{f}$: $\bm{f}(t)\!=\!\{f^r(t)| r \!\in\! \mathcal{C}_{u}(t) \}$, where $f^r(t)$ indicates whether demand $r$ arrives at destination BS $b \!\in\! \mathcal{B}$, defined as
    \begin{equation}
        f^r{(t)}\!=\!
        \begin{cases}
        \!1,  \text{if }a_u^r{(t)} \text{ is the destination BS }b, \\
        \!0,  \text{otherwise.} 
        \end{cases}
    \end{equation}
    \item Discount factor $\bm{\gamma}$: $\bm{\gamma}=\{\gamma_u| u\!\in\! \mathcal{U}\}$, 
    in which $\gamma_u$ is designed to calculate the cumulative reward. 
    A larger $\gamma$ indicates decisions focusing on the long-term reward.
\end{itemize}

% The agents interact with the network environment to learn a reward-maximum policy. In this paper, 
% cooperation among the agents is achieved either via the communication between neighboring nodes. 
% In the next section, we will first introduce the framework of the RL algorithm.
The policy $\pi_{u}^r(t)$ leads agent $u$ to select action $a_{u}^r(t)$ for demand $r$
under observation $o_{u}(t)$ at time step $t$.
$\bm \pi_u(t)=\{\pi_{u}^r(t), \forall r \!\in\! \mathcal{C}_u(t)\}$ indicates the joint policy of all demands on agent $u$. 
$\bm \Pi(t)=\{\bm\pi_{u}(t), \forall u \in \mathcal{U}\}$ indicates the policy of all agents. 
According to the Dec-POMDP and specific policy $\bm \Pi(t)$, 
we can obtain the routing path for all demands. 
Therefore, $\mathscr{P}0$ is transformed to find the optimal 
policy $\bm \Pi^{\star}(t)=\{\bm\pi_{u}^{\star}(t), \forall u\!\in\! \mathcal{U}\}$, 
and then the improved MADRL-based routing method is designed to 
obtain $\bm \Pi^{\star} =\{\bm \Pi^{\star} (t)| t \in T\}$  for minimizing the total E2E delay.
\subsection{SHERB-MADDQN-based Routing Algorithm}
% In the multiple sources and multiple destinations dynamic scenario, 
% to improve the learning efficiency of routing strategies, the adaptive SHERB-MADDQN-based routing algorithm is designed.  
In dynamic scenarios with multiple sources and destinations,
to improve the learning efficiency of routing strategies, the adaptive SHERB-MADDQN-based routing algorithm is designed.  
In detail, the experience buffer of each agent is softly constructed 
by embedding the state information of the destination BS in the reward.
% To effectively deal with the complex issue of multiple destination routing in UAV networks,
Thus, the reward in ({\ref{equ:reward}}) is reformulated as 
\begin{equation}
    \begin{aligned}
    \mathcal{R}_{u}^r(t)=\frac{1}{\mathcal{T}^{tr,t}_{u\kappa}(t)+\varsigma} \cdot& \frac{d_{u\kappa}(t)}{d_{u\kappa}(t)+d_{\kappa b}(t)} \eta_{u \kappa}^r(t),\\
    &r \in  \mathcal{C}_{u}(t), \kappa \in \Gamma_{u}(t), t\in T,
    \end{aligned}
\end{equation}
where $d_{u\kappa}(t)$ and $d_{\kappa b}(t)$ denote distances
from UAV $u$ to neighboring UAV $\kappa$, and from UAV $\kappa$ to destination BS $b$, respectively.
$\varsigma$ is a hyperparameter to balance the benefits from the delay and  distance.
It is noted that $d_{u\kappa}(t)$ is positively related to the reward, 
since a larger value of $d_{u\kappa}(t)$ indicates a fewer hop counts required, 
saving time in queuing for transmission.
Meanwhile, ${d_{u\kappa}(t)+d_{\kappa b}(t)}$ represents the total transmission distance 
from the selected next hop $\kappa$ to UAV $u$ and BS $b$.
If $d_{u\kappa}(t)+d_{\kappa b}(t)$ is closer to the straight shortest distance between UAV $u$ and BS $b$, 
agent $u$ may obtain the greater value of rewards, and vice versa.
% It enables the classification learning of demands with different destinations, 
Moreover, compared with the hard-HERB (HHERB)-based method that constructs 
a hierarchical data structure to store the experience, 
the SHERB can mitigate the problem of sparse experience samples, 
improving the sample utilization efficiency and generalization ability of strategies.
% In the paper, ERB each agent $u$ maintains a 
% $\mathcal{D}_u=\{( o_{u}(t), a_{u}^r(t), \mathcal{R}_{u}^r({t+1}), o_{u}({t+1}))\vert r\in \mathcal{C}_u(t)\}$.

Furthermore, when the dimensions of state and action spaces are large, it is tricky to maintain the Q-table in Q-learning. 
Besides, the deep Q-network (DQN)-based algorithm may cause a large deviation  
caused by overestimating the Q-target value. 
Hence, by combining double Q-learning with DQNs, DDQN algorithms are proposed to approximate 
Q-value functions via deep neural networks (DNNs) and decouple the action selection 
and calculation of Q-target values {\cite{van2016deep}}. 
% It is theoretically proved the DDQN-based method can avoid the overestimation .
% Besides, agent $u$ equip each $\mathcal{D}_u$ with an independent DDQN network,
% which allows each network to focus on the value function fitting for specific demands, 
% avoiding policy interference. 
% The network takes the observation of the agent $u$ and the destination node $b$ of the demand as inputs, 
% and outputs the Q-value of the action corresponding to this destination node.
Further, for the DDQN, there exist two DNNs, i.e., 
the online network with parameter $\theta_{u}$ 
and the target network with parameter $\theta^{-}_{u}$ of each agent $u$.
Through constantly updating the weight $\theta_{u}$ of online networks,
loss function $L(\theta_{u})$ is trained and then minimized. 
The experience replay buffer (ERB) of agent $u$ is designed to store the historical experience and denoted as 
$\mathcal{D}_u\!=\!\{(o_{u}(t), a_{u}^r(t), \mathcal{R}_{u}^r({t+1}), o_{u}({t+1}))\vert r \!\in\! \mathcal{C}_u(t)\}$.
By randomly sampling $D$ transition tuples from $\mathcal{D}_u$,
$L(\theta_{u})$ is typically computed as the mean squared error between Q-value function
$Q(o_{u}(t), a_{u}(t);  \theta_{u})$ and Q-target $y_{u}(t)$, i.e.,
\begin{algorithm}[t!]
    \caption{{\label{algorithm-MADDQN}}SHERB-MADDQN-based routing algorithm}
    \begin{algorithmic}[1]
    \REQUIRE {$\mathcal{I}$, $\mathcal{U}$, $\mathcal{B}$, $\mathcal{E}$, $R$, ${\bm{\alpha}}$, and ${\bm{\gamma}}$ }.
    \ENSURE Optimal policy ${\bm{\Pi^{*}}}$.
    \STATE\textbf{Initialization:}\label{initialize} Initialize the network environment, hyper-parameters, 
    ERB set $\bm {\mathcal{D}}$, and the online 
    and target network parameters $ \theta_u $  and $ \theta_u^{-} $ for each agent $u \in \mathcal{U}$, respectively.
    \FOR{each episode} {\label{episode}}
    \FOR{$t = 1, \dots, T$}
    \FOR{$u = 1,\dots, U$} % 遍历每个无人机智能体
    \STATE {\label{episode-initi}} The observation of agent $u$ is set as $o_{u}(t)$.
    \STATE {\label{Allocate-BW}} Allocate the bandwidth based on queued demands in $\mathcal{C}_u(t)$ via (\ref{equ:BW}). 
    \FOR{$r = 1, \dots, |\mathcal{C}_u(t)|$}{\label{FOR-demands}} % 遍历无人机$j$携带的每个数据包
        \STATE Select action $a_{u}^r(t)$ for demand $r$ under observation $o_{u}(t)$ 
        using an $\epsilon$-greedy policy.
        \STATE Execute $a_{u}^r(t)$, and obtain reward $\mathcal{R}_{u}^r(t\!+\!1)$.
    \ENDFOR{\label{ENDFOR-demands}}
   
     \IF{$|\mathcal{D}_u| > D$} {\label{sample}}
        \STATE Randomly select $D$ samples from $\mathcal{D}_u$.
        \STATE {\label{target}} Compute Q-target value $y_{u}(t)$ via (\ref{Q-target}).
        \STATE {\label{update-theta}} Update $\theta_u$ via the gradient descent in (\ref{theta-update}). 
        \STATE {\label{update-theta-}} Periodically update target network parameter $\theta_u^{-}$ via (\ref{target-undate}) every $\mathcal{W}$ steps.
    \ENDIF 
    \ENDFOR
    \STATE {\label{update-environment}} Update the environment, 
    and set observation $o_{u}(t)\leftarrow o_{u}({t + 1})$ for all agents in $\mathcal{U}$.
    \STATE {\label{Store-E}}Store transition $( o_{u}(t), a_{u}^r(t), \mathcal{R}_{u}^r(t\!+\!1), o_{u }({t\! + \!1}))$.
     of each demand $r$ into corresponding $\mathcal{D}_u$.
    % \STATE {\label{update-observation}} .
    \ENDFOR
    \ENDFOR
    \end{algorithmic}
\end{algorithm}
\begin{equation}{\label{loss}}
    L(\theta_{u}) = \mathbb{E}_{\xi \thicksim \mathcal{D}_{u}} \left[ \left( y_{u}(t)  
    - Q(o_{u}(t), a_{u}(t);  \theta_{u}) \right)^2 \right],
\end{equation}  
where tuple $\xi$ is a transition data of $\mathcal{D}_{u}$, and $y_{u}(t)$ is
\begin{equation}{\label{Q-target}}
    \begin{aligned}
         &y_{u}(t)  =  \mathcal{R}_{u}({t+1}) +  \\
          &\gamma_u Q(o_{u}({t+1}), \arg\max_{a^{\prime}_{u}(t)} Q(o_{u}({t+1}), a^{\prime}_{u}(t);  \theta_{u}^b)|\theta^{-}_{u}).
    \end{aligned}
\end{equation}
Besides, the gradient of $L(\theta_{u})$ is denoted as $\nabla_{\theta_{u}}L(\theta_{u})$,   
which is used to  update $\theta_{u}$ via the gradient descent, i.e.,
\begin{equation}{\label{theta-update}}
    \theta_{u} \leftarrow \theta_{u} - \alpha_{u} \nabla_{\theta_{u}} L(\theta_{u}).
\end{equation}
Wherein, $\alpha_{u}$ is the learning rate that controls the size of update steps.
In every $\mathcal{W}$ steps, parameter $\theta^{-}_{u}$ is periodically updated 
to match the parameter of online network $\theta_{u}$ for stabilizing training 
and improving the convergence, i.e.,
\begin{equation}{\label{target-undate}}
    % \theta_{i}^{-}\leftarrow \ell\cdot \theta_{i}+ (1-\ell)\cdot \theta_{i}^{-}.
    \theta_{u}^{-}\leftarrow \tau \theta_{u} +(1-\tau)\theta_{u}^{-},
\end{equation}
where $\tau$ is the soft update coefficient.

The detail of the proposed SHERB-MADDQN-based routing method is shown in Algorithm {\ref{algorithm-MADDQN}}.
Firstly, the algorithm initializes the network environment,
hyper-parameters, ERB set $\bm{\mathcal{D}}$, and the parameters of DNNs, respectively ({line \ref{initialize}}).
At the beginning of each episode, the observation of agent $u$ is initialized as $o_{u}(t)$ (lines {\ref{episode}}-{\ref{episode-initi}}).
Then, according to (\ref{equ:BW}), the bandwidth of agent $u$ is allocated according to queued demands in $\mathcal{C}_u(t)$ (line \ref{Allocate-BW}).
Based on $o_{u}(t)$, each agent $u$ selects sub-action $a_{u}^r(t)$ for demand $r$ via an $\epsilon$-greedy policy, 
and then executes $a_{u}^r(t)$ to obtain reward $\mathcal{R}_{u}^r({t+1})$ (lines {\ref{FOR-demands}}-{\ref{ENDFOR-demands}}). 
If the number of transition tuples in $\mathcal{D}_{u}$ is larger than mini-batch $D$, $D$ samples are randomly selected to calculate 
Q-target value $y_{u}(t)$ via (\ref{Q-target}) (lines {\ref{sample}}-{\ref{target}}). 
Additionally, the DNN performs a gradient descent step to update $\theta_{u}$ (line {\ref{update-theta}}).
The target network parameter $\theta_{u}^{-}$ is periodically updated according to ({\ref{target-undate}}) 
every $\mathcal{W}$ steps (line {\ref{update-theta-}}).
At time step $t$, when all agents finish the actions of all demands, the environment is updated.
Further, each agent $u \!\in \! \mathcal{U}$ obtains next observation $o_{u}({t \!+ \!1})$ to update observation $o_{u}(t)$ (line \ref{update-environment}).
Besides, the transition $( o_{u}(t), a_{u}^r(t), \mathcal{R}_{u}^r(t\!+\!1), o_{u }({t\! + \!1}))$ 
of each demand $r$ is stored into corresponding $\mathcal{D}_u$ for training (line \ref{Store-E}).
% Set $o_{u}(t)\leftarrow o_{u}({t + 1})$ for all agents in $\mathcal{U}$. 
% Besides, the observation $o_{u}(t)$ of agent $u$ is correspondingly updated by $o_{u }({t+1})$ (line \ref{update-observation}).

\section{Simulation Results\label{sec:Simulation Results}}
In this section,  a couple of simulations are conducted via Python. 
% The specific parameters are listed in Table {\ref{table1}} {\cite{channel_model_PL}}. 
Specifically, nodes of LAINs are distributed in the area within a 15km $\times$ 5km range, 
the altitude range of UAVs is within [0.2, 0.4]km. 
UAVs are initialized by the ground control center at the beginning of  missions.
Besides, the size of demands from SDs is randomly set within [400, 600]kbits.
Other parameters are set as: $d_{min}\!=\!$ 10m,  $d_{u,max}\!=\!$ 100m, 
$\sigma_{ij}^{2}\!=\!-$110dBm, $\lambda\!=\!$ 2.4Ghz, $c\!=\!3 \times 10^{8}$m/s, $P^{tr}\!=\!$ 40dBm, 
$C^{max}_{i}\!=\!$ 50, $B_{u}\!=$ 2MHz,
$\varrho_1\!=\!$ 5.0188, $\varrho_2\!=\!$ 0.3511, 
$\eta_{n m}^\mathrm{LoS}\!=\!$ 0.1dB, and
$\eta_{n m}^\mathrm{LoS}\!=\!$ 21dB {\cite{JiaHiera,}}.

% Fig.{\ref{fig:Delay}}

% Fig.{\ref{fig:Step}}

\begin{figure}[t]
    \vspace{-0.3cm}
    \centering
    \includegraphics[width=0.7\linewidth]{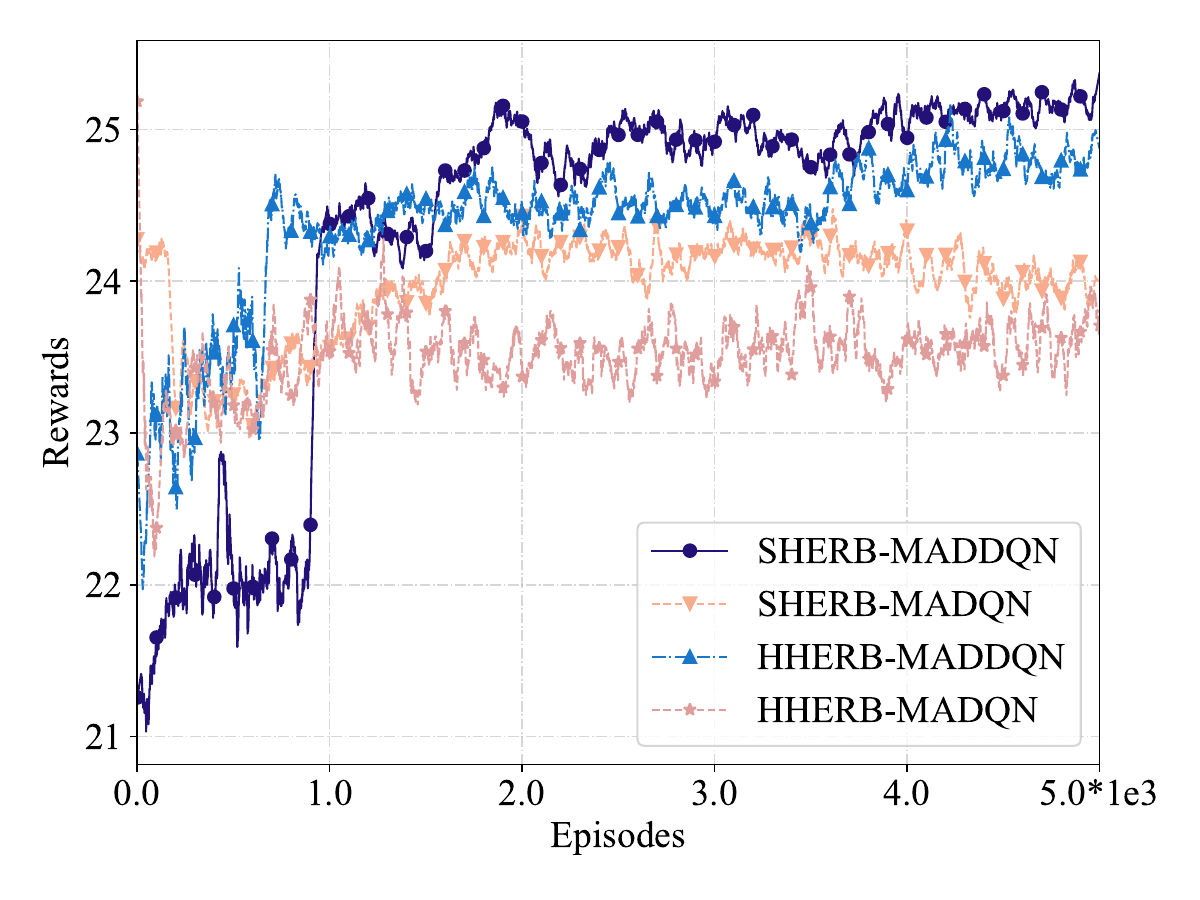}
 
        \caption{\label{fig:Rewards}Comparisons of the four algorithms in terms of the accumulative reward during the training process. 
       \vspace{-0.3cm}
    } 
    
\end{figure}
\begin{figure}[t]
    \centering
    \includegraphics[width=0.7\linewidth]{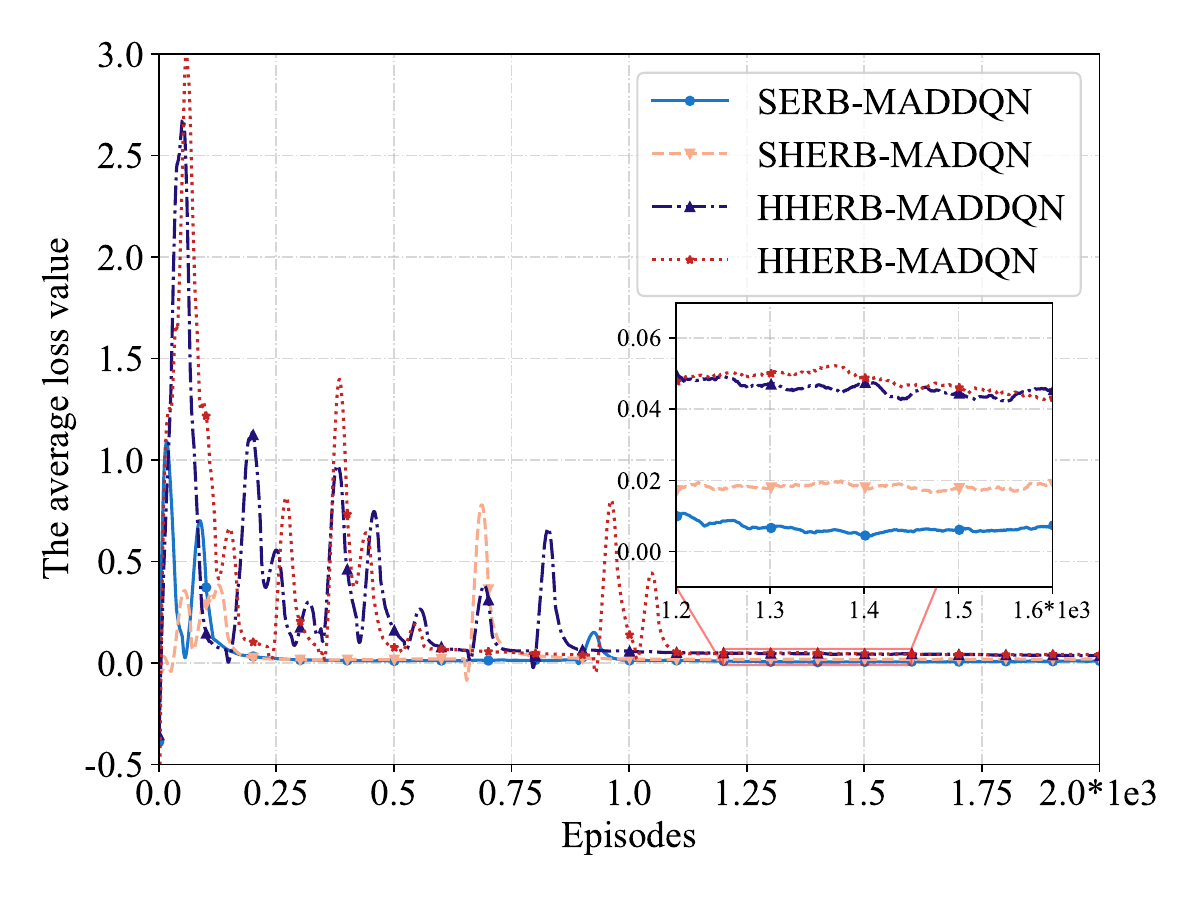}
    \caption{\label{fig:Loss}Comparisons of the four algorithms in terms of the convergence performance during the training process. 
    } 
    \vspace{-0.3cm}
\end{figure}

\begin{figure}[t]
    \centering
            \vspace{-0.3cm}
    \includegraphics[width=0.7\linewidth]{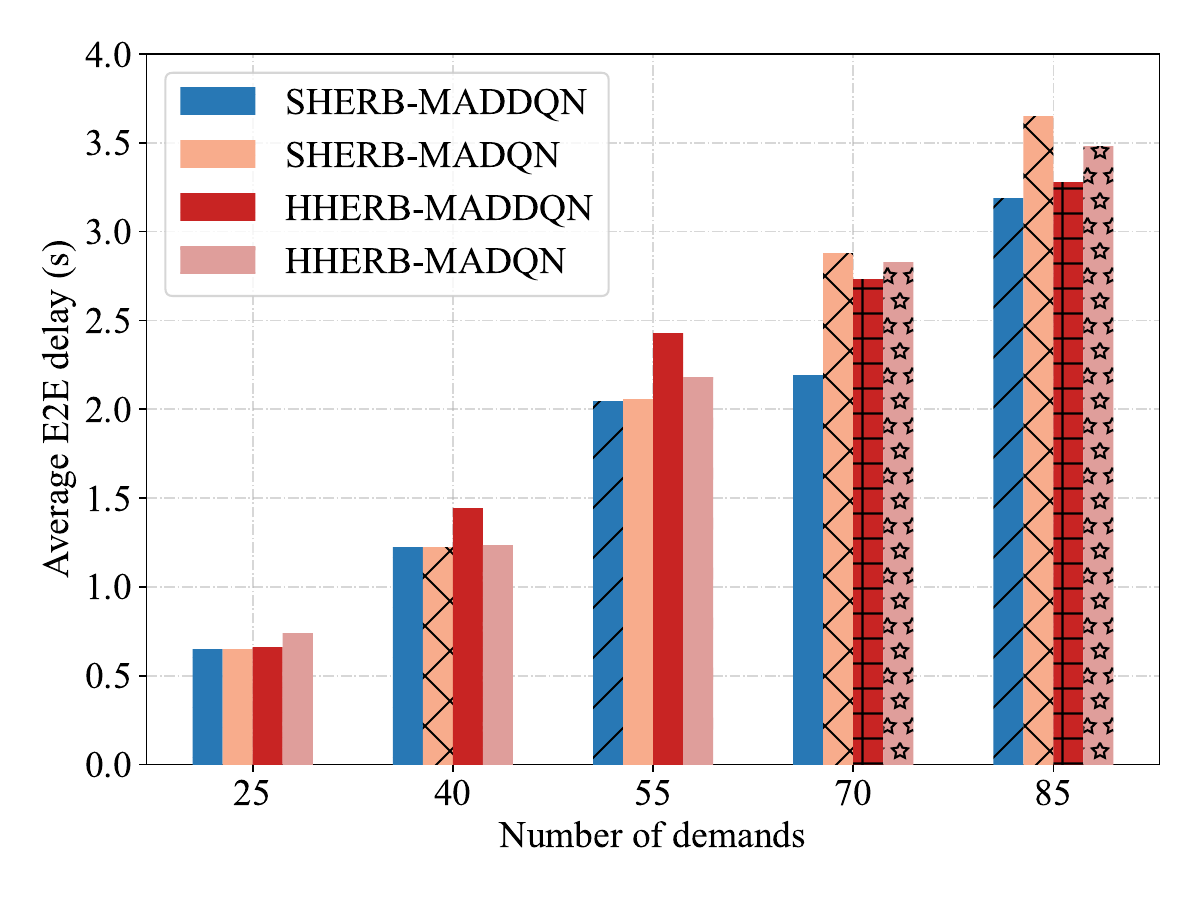}
    \caption{\label{fig:Delay}Comparisons of the four algorithms in terms of average E2E delay. 
    } 
        \vspace{-0.3cm}
\end{figure}

\begin{figure}[t]
    \centering
    \includegraphics[width=0.7\linewidth]{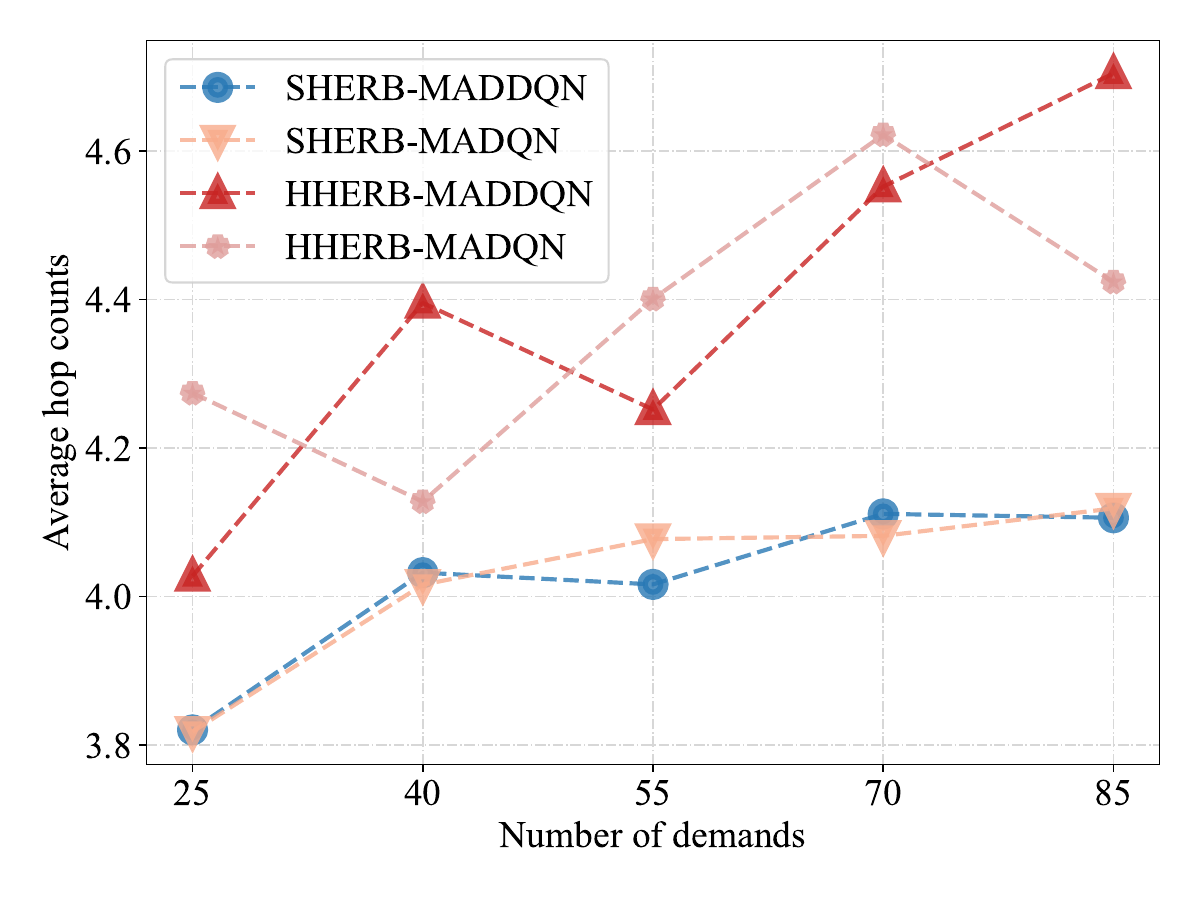}
    \caption{\label{fig:Step}Comparisons of the four algorithms in terms of  average hop counts. 
    } 
        \vspace{-0.3cm}
\end{figure}

% \begin{figure*}[t]
%     \vspace{-0.2cm}
%     \centering
%     % 设置子图说明居中
%     % \captionsetup[subfloat]{justification=centering,singlelinecheck=false}
%     % % 设置大图说明左对齐
%     % \captionsetup[figure]{justification=raggedright,singlelinecheck=false}
%     \subfloat[\textcolor{black}{ }]{\centering
%     \includegraphics[width=0.33\linewidth]{Loss.pdf}}
%     % \hspace{0.01\linewidth}
%     \subfloat[\textcolor{black}{}]{\centering
%     \includegraphics[width=0.33\linewidth]{Delay.pdf}}
%     \subfloat[\textcolor{black}{}]{\centering
%     \includegraphics[width=0.33\linewidth]{Step.pdf}}
%     \caption{Comparisons of the four algorithms with different metrics. 
%     (a) Convergence performances of the average loss value. 
%     (b) Average E2E delay with different demand numbers.
%     (c) Average hop counts with different demand numbers.}
%     \label{fig:lr-number-demand}
%     \vspace{-0.4cm}
% \end{figure*}

To demonstrate the performance of Algorithm 1, different metrics are evaluated 
by comparing with SHERB-MADQN, HHERB-MADDQN, and HHERB-MADQN algorithms in Figs. {\ref{fig:Rewards}}-{\ref{fig:Step}}.
Specifically, in Fig. {\ref{fig:Rewards}}, the accumulative rewards are provided to evaluate the convergence performance.
It can be observed that all four algorithms converge with different performances.
In the first 2,000 episodes, rewards are not satisfactory. 
As the number of episodes increases, rewards increase and converge at the specific values.
Besides, the rewards of SHERB-MADRL-based methods are larger compared to the rewards of HHERB-MADRL-based algorithms, 
due to the higher learning efficiency of the proposed SHERB.
Meanwhile, the convergence value of the proposed SHERB-MADDQN algorithm is larger than the SHERB-MADQN algorithm. 
Hence, according to the designed (\ref{equ:reward}) where rewards are negative, the proposed algorithm may have less delay for routing. 
% In Fig. {\ref{fig:lr-number-demand}}, the SHERB-MADDQN is compared with 
% other three algorithms with regard to the convergence performance of training losses, average E2E delay, and rouitng hops. 
% Specifically, in Fig. {\ref{fig:lr-number-demand}}(a), the loss functions of all four algorithms converge.
% For the proposed SHERB-MADDQN, the convergence speed is faster and the curve fluctuation of loss functions has a narrower range.
% Moreover, the loss value of the proposed method is smaller compared to other algorithms when the curve converges, 
% indicating a more stable and effective training process.
% The E2E delay for transmitting demands is evaluated in Fig. {\ref{fig:lr-number-demand}}(b).
% It is observed that as the demands grow, the E2E delay of the data transmission 
% increases of all algorithms, due to the limited resources of the bandwidth. 
% Nevertheless, the delay of the SHERB-MADDQN algorithm decreases by 24.09\%, 20.49\%, and 22.56\% 
% than the delay of SHERB-MADQN, HHERB-MADDQN, and HHERB-MADQN algorithms, respectively.
% Fig. {\ref{fig:lr-number-demand}}(c) shows the performance of hop counts and the demand numbers. 
% Specifically, with the increasing demands, the average hop counts of all algorithms grow, 
% due to the increased congestion of networks.
% Nevertheless, the SHERB-based algorithm requires fewer hop counts for transmitting demands, 
% benefiting from the better algorithm performance.
% In short, simulation results demonstrate that the proposed algorithm performs better during routing.
In Fig. {\ref{fig:Loss}}, the loss functions of all four algorithms converge.
For the proposed SHERB-MADDQN, the convergence speed is faster and the curve fluctuation of loss functions has a narrower range.
Moreover, the loss value of the proposed method is smaller compared to other algorithms when the curve converges, 
indicating a more stable and effective training process.

The E2E delay for transmitting demands is evaluated in Fig. {\ref{fig:Delay}}.
It is observed that as demands grow, the E2E delay of the data transmission 
increases  for all algorithms, due to the limited resources of the bandwidth. 
Nevertheless, the delay of the SHERB-MADDQN algorithm decreases by 24.09\%, 20.49\%, and 22.56\% 
than the delay of SHERB-MADQN, HHERB-MADDQN, and HHERB-MADQN algorithms, respectively.
Fig. {\ref{fig:Step}} shows the performance of hop counts under different demand numbers. 
Specifically, with the increasing demands, the average hop counts of all algorithms grow, 
due to the increased congestion of networks.
Nevertheless, the SHERB-based algorithm requires fewer hop counts for transmitting demands, 
benefiting from the better training performance of the algorithm.
In short, simulation results demonstrate that the proposed algorithm performs better during routing.
% 9.1、13.2、11.18。
% 第一组数据相较于第 2 组数据的增降百分比: [0.0, 0.0, -0.54, -24.09, -12.5]
% 第一组数据相较于第 3 组数据的增降百分比: [-1.52, -15.45, -15.79, -20.49, -2.72]
% 第一组数据相较于第 4 组数据的增降百分比: [-12.19, -1.22, -6.32, -22.56, -8.28]
\section{Conclusions\label{sec:Conclusions}}
In this work, we characterize the routing process of multiple sources and destinations in zero-trust LAINs, with the joining and exiting of UAVs.
Additionally, the blockchain technique is introduced to manage the mobility and verify identities, 
enhancing the reliability and safety during routing. 
Besides, the routing problem is formulated to minimize the total E2E delay with multi-constraints. 
Further,  we reformulate the routing problem into a Dec-POMDP to deal with the challenge of obtaining global information in LAINs 
characterized by the high dynamic, distributed topology.
Then, to improve the learning efficiency of the MARL, a SHERB-MADDQN-based algorithm is proposed via 
embedding the observed state into rewards. 
Simulation results in LAINs demonstrate that the designed SHERB-MADDQN algorithm outperforms 
in the delay, convergence, and loss value of training performances than other algorithms.

\bibliographystyle{IEEEtran}
\bibliography{ref2}

\end{document}